\documentclass[aps, jcp, prb,  two column,showpacs,groupedaddress]{revtex4-1}
\usepackage{float}
\usepackage{graphicx} 
\usepackage{subfigure}
\usepackage{amsmath}
\usepackage{dcolumn}
\usepackage{multirow}
\usepackage{bm}     
\usepackage{amssymb,amssymb}   
\usepackage[left= 1 in, right= 1 in, top= 1 in , bottom= 1 in]{geometry}
\usepackage{hyperref}
\usepackage{xcolor} 
\hypersetup{ colorlinks, citecolor=blue ,linkcolor=black }

\usepackage{chemfig}

\usepackage{hyperref}

\hypersetup{linktocpage}

\begin{document}
	\vspace*{0.35in}

	\title{ Transport properties of Gamma-Aminobutyric Acid in water}
	
	\author{Esha Mishra}
  
	\author{ Narayan Prasad Adhikari }
	\email{npadhikari@tucdp.edu.np}
	\affiliation{Central Department of Physics, Tribhuvan University, Kirtipur, Kathmandu, Nepal}

	\begin{abstract}
		\noindent In the present work, molecular dynamics study of diffusion of Gamma-Aminobutyric acid in water at different temperatures (298.2 K, 303.2 K, 313.2 K, 323.2 K, 333.2 K) have been performed. The solute and solvent is modeled using OPLS-AA platform. The structure of the system is analyzed using RDF of different atom/molecule pairs. In all the cases, at least two or more peaks were obtained which suggests appreciable amount of interaction between atoms and molecules. The self diffusion coefficient of GABA and water is calculated from mean square displacement (MSD) plot using Einstein's relation. Binary diffusion coefficient is obtained from self diffusion coefficient using Darken's relation. Binary diffusion coefficient agrees well with the previously reported experimental data within an error of around 7\%. The temperature dependence of self diffusion coefficients of GABA in water follows the Arrhenius behaviour. With the help of Arrhenius plot, we estimate the activation energy. The activation energy of water agrees well with the previously reported experimental data within an error of around 17\%.\\
		
	\end{abstract}
	
	\maketitle
	
	\section*{Introduction}
	\noindent Amino acids are regarded as the building block of proteins. An amino acid contains an amino group ($-$NH$_2$) and carboxyl group ($-$COOH) as its functional group along with an alkyl group (-R) as a side chain which determines the specific type of amino acids.  
	Gamma-Aminobutyric Acid ($\gamma$-Aminobutyric acid) is a four carbon, non protein amino acid having $\gamma$ carbon linked with an amino group~\cite{boyd}. It is also known as GABA. The IUPAC name of GABA is 4-aminobutanoic acid. GABA is highly soluble in water. It is found mostly as a zwitterion such that the carboxyl group is deprotonated and the amino group is protonated. The chemical formula of GABA is C$_4$H$_9$NO$_2$.
	
	\noindent GABA is synthesized in our body from glutamate in the presence of enzyme L-glutamic acid and decarboxylase. These enzymes convert glutamate into gamma-aminobutyric acid~\cite{enz}. 
	The deficiency of GABA causes anxiety, drowsiness, blurred vision, irritation, memory problems, fatigue and swelling of limbs. GABA is also used as a drug for anti-anxiety and anticonvulsive effects. With the help of diffusion phenomena, the tuning of inhibitory neurotransmission of GABA receptor is performed. The results of the experiment concluded that the GABA diffusion is linked to the changes in the intracellular $Ca^{2+}$ concentration~\cite{gabatun}. The system of GABA transporters in neural cell plays an efficient role in terminating the GABAergic neurotransmission~\cite{inhib}. Experiments have been carried out to demonstrate the involvement of GABA transport in hepatic uptake of taurine in rats~\cite{rat}. The GABA and Glutamate transporters are reported to keep the excitatory amino acids low and provide amino acids for the metabolic purposes~\cite{iv}. We can conclude that the GABA transport plays a significant role in the mammalian central nervous system.\\
	
	\noindent Molecular dynamics simulations have been performed to study the solvation effects of water and trifluoroethanol on gamma-aminobutyric acid. On studying the GABA-water interactions, the simulations yielded results that showed strong interaction among the GABA molecules and water molecules. This explained that GABA is capable of breaking water-hydrogen bonds and interfering in the hydrogen bonding 
	of the corresponding water~\cite{gabasolv}. Water is an indispensable part of biological macromolecules. Life on earth cannot be imagined without water. Water holds a huge significance in all the metabolic processes taking place in our body. Despite the experimental estimation of the diffusion process of GABA~\cite{kyui}, to the best of our knowledge, there has been no molecular dynamics study on the diffusion of GABA in water.\\
	\section*{Diffusion}
	\noindent Diffusion is the process by which matter is transported from one region of a system having higher concentration to the region of a system having lower concentration as a result of random molecular motions~\cite{crank}. Self diffusion is the diffusion process which occurs in a homogeneous system where no chemical concentration gradient exists. It is measured in terms of self diffusion coefficient. The mathematical expression used to calculate the self diffusion coefficient from molecular position is known as the \textit{Einstein's relation}~\cite{ipna}. For a 3D system,
	the relation between the self diffusion coefficient and mean square displacement is given by:
	\begin{equation}
	\lim_{t \to \infty} \langle \lbrack r(t) - r(0) \rbrack ^2 \rangle = 6Dt \label{selfdiff}
	\end{equation}
	
	Here, $r(t) - r(0)$ is change of position  of diffusing particle in time t. $\langle .... \rangle$ represents the ensemble average of quantity inside the angled bracket. The mean square displacement of diffusing particle is given by $\langle \lbrack r(t) - r(0) \rbrack ^2 \rangle$. The graph with time along x - axis and mean square displacement along y - axis gives mean square displacement plot. The slope of the MSD plot when divided by six gives self - diffusion coefficients.\\

	\noindent Binary diffusion is the diffusion process which occurs in a heterogeneous system containing two different species in a binary mixture. It is measured in terms of  diffusion coefficient. The expression for binary diffusion coefficient is given by \textit{Darken's relation} ~\cite{darken}.

	\begin{equation}
	D_{12} = N_2D_1 +N_1D_2 
	\end{equation}

	where,  
	$D_{12}$ = binary diffusion coefficient,
	
	$D_1, D_2$ = self diffusion coefficients of species 1 and 2 respectively,

	$N_1, N_2$ = mole fractions of species 1 and 2 respectively.

	\section*{Computational Details}
	\subsection*{Modeling of System}
	\noindent The method used for computing the equilibrium and transport properties of classical many body systems is molecular dynamics. It provides detailed microscopic modeling on the atomic or molecular scale to study the dynamic of the molecule. If the positions and velocities of an atom are known, molecular dynamics is one of the best methods to predict the state of a system at any time. Also, the trajectory helps to calculate the equilibrium and the transport properties~\cite{frenkel}. In MD simulations, we proceed by solving Newton's equation of motion for a system of N atoms interacting according to a potential energy $U$.
	\begin{equation}
	\label{Newton'sequation}
	m_i\frac{\partial^2\textbf{r}_i}{\partial t^2} = -\nabla_i U(r) =\textbf{F}_i
	\end{equation}
	where $m_i$ is mass of $i^{th}$ particle, $r_i$ is the position of the particle, U(r) is the average potential experienced by the i$^{th}$ particle and $\textbf{F}_i$ is the mean force on the particle. \\

	\noindent We begin the process of molecular dynamics by preparing the model of the system under our research problem. The potential functions for the interaction of these atoms and molecules are derived empirically. The instantaneous force acting on an individual particles is calculated using the potential functions along  with the force field parameters~\cite{allen}. The total potential energy of a system is sum of all interaction potential energy:
	\begin{equation}
	\label{totalpotential}
	U_{total} = U_{bond} + U_{angle} + U_{proper} + U_{LJ} + U_{Coulomb}
	\end{equation}

	\subsection*{Simulation Set Up} 
	
	\noindent Simulation was carried out for GABA-water system at a different temperatures. GABA is a four carbon, non protein amino acid. The amino group is linked with the gamma carbon. The different atoms in the molecule have different partial charges due to the difference in electronegativity. The electrostatic properties are determined by the partial charges. The LJ parameters define the van der Waals interaction. The bonded and non-bonded interactions are also taken into account. The interactions were parametrized by using OPLS-AA force field.  We have used SPC/E model of water in our simulation. The water model maintains an electrically neutral water molecule by taking 3 atoms (2 H-atoms and 1 O-atom) at each atomic site with a partial charge of +0.4238e for H atom and $-0.8476e$ for O- atom where $e$ is the elementary charge having magnitude $1.6022 \times 10^{-19}$ Coulomb ~\cite{npna}.\\
	
	\noindent In GROMACS, these parameters are presented in file named \textit{spce.itp}.
	The force field parameters for flexible SPC/E model are presented below:
	
	\begin{table} [H]
		\centering
		\label{forcefieldforH2O}
		\caption{Force field parameters for SPC/E water model.}
		\begin{tabular}{|c|c|} \hline
			
			 Parameters &Values\\
			  \hline
			$K_{OH}$ & $ 3.45 \ \mathrm{x} \ 10^5 \ \mathrm{kJmol^{-1}nm^{-2}} $ \\
			\hline
			$b_{OH} $ & $ 0.1 \mathrm{nm}$ \\
			\hline
			$K_{HOH}$ & $3.83 \  \mathrm{x} \ 10^2  \ \mathrm{kJmol^{-1}rad^{-2}}   $\\
			\hline
			$\Theta_o $ & $ 109.47 \ \mathrm{deg}$ \\
			\hline
		\end{tabular}
	\end{table}	
	
	\noindent The information about the non bonded interaction is specified in file \textit{ffnonbonded.itp} which is presented below:\\
	
	\resizebox{0.45\textwidth}{!}{\begin{tabular}{ccccccc}
			[\textbf{atomtypes}]&& &&&&\\
			
			name &  at.num &   mass   &     charge &  ptype &        c6     &      c12\\
			H1 &    1   &   1.0080   &    0.060  &   A  &  2.50000e-01     & 6.27600e-02 \\
			H2 &    1   &   1.0080   &    0.060  &   A  &  2.50000e-01     & 6.27600e-02 \\
			H3 &    1   &   1.0080   &    0.060  &   A  &  2.50000e-01     & 1.25520e-01 \\
			H4 &    1   &   1.0080   &    0.060  &   A  &  2.50000e-01     & 1.25520e-01 \\
			H5 &    1   &   1.0080   &    0.060  &   A  &  2.50000e-01     & 1.25520e-01 \\
			H6 &    1   &   1.0080   &    0.060  &   A  &  2.50000e-01     & 1.25520e-01 \\
			H7 &    1   &   1.0080   &    0.450 &   A  &  0.0000e+00     & 0.0000e+00 \\
			H8 &    1   &   1.0080   &    0.360 &   A  &  0.0000e+00     & 0.0000e+00 \\
			H9 &    1   &   1.0080   &    0.360 &   A  &  0.0000e+00     & 0.0000e+00 \\
			N1 &    1   &   14.00670   &  -0.900 &   A  &  3.30000e-01     & 7.11280e-01 \\
			
			C1 &    6   &  12.01100  &   0.060  &   A  &  3.50000e-01  &  2.76144e-01    \\ 
			C2 &    6   &  12.01100  &   -0.120  &   A  &  3.50000e-01  &  2.76144e-01    \\ 
			C3 &    6   &  12.01100  &   -0.120  &   A  &  3.50000e-01  &  2.76144e-01    \\ 
			C4 &    6   &  12.01100  &   0.520  &   A  &  3.75000e-01  &  4.39320e-01    \\  
			O1 &      8 &   15.99940 &     -0.440 &    A & 2.96000e-01 & 8.78640e-01\\  
			O2 &      8 &   15.99940 &     -0.530 &    A & 3.00000e-01 & 7.11280e-01\\  
			
		\end{tabular}}\\
		
		\noindent First column represents the name of atoms. The second column presents the atomic number of a respective atom and the third column gives the mass of atoms in the atomic mass unit. The fourth column presents partial charge in atoms. The fifth column gives the particle type and here A stands for the atom. Sixth and seventh columns give parameters C6 and C12 respectively of the corresponding atom.\\
		
\begin{figure}
			\centering
			\includegraphics[scale=0.3]{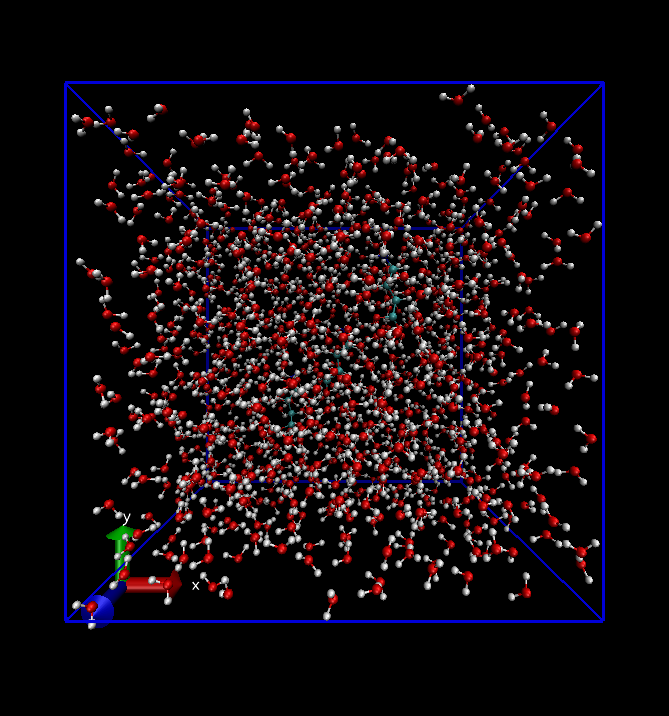}
			\caption{System under study after energy minimization.}
			\label{fig:enm box}
		\end{figure}

		\noindent Before solvation, only three molecules of GABA are present in the box.
		A complete system for our simulation, after the solvation of molecules in water.
		The process of energy minimization is employed in order to ensure that our system for simulation is not far from equilibration. We have used steepest descent method for energy minimization which is represented by $integrator = steep$. $emtol =50.0$ represents force tolerance in units $kJmole^{-1}nm^{-1}$, $emstep = 0.002$ gives the step size for position in $nm$.  If no particle in the system experiences forces greater than \textit{emtol}, the system is said to be in the state of minimum energy. Temperature and pressure coupling is not required in this step. The system after adding 1035 water molecules in simulation box containing 3 GABA molecules is shown in FIG.\ref{fig:enm box} .\\ 
		
		\noindent After energy minimization, the system is in one of the local potential minima. In GROMACS, the system is brought in thermal equilibrium by coupling it with thermostat and barostat. Thermostat and barostat are used to rescale temperature and pressure of the system to the desired value. Rescaling temperature using thermostat is also termed \textit{temperature coupling} and rescaling pressure of system using barostat is termed as \textit{pressure coupling}. Particle Mesh Ewald (PME)
		is a method employed for the better performance of the reciprocal sum. This method assigns a charge to the grid using interpolation. The grid undergoes Fourier transformation which yields reciprocal energy term through a single sum over the grid in K - space. The inverse transformation is used for the calculation of the potential at grid points. The forces on each atom are obtained by using the interpolation factors. It is used for coulomb interaction with Fourier spacing of 0.3 with cutoff distance 1.0 nm~\cite{manual}. Velocity -rescaling is used for temperature coupling. Our run is carried out at five different temperatures 298.2 K, 303.2 K, 313.2 K, 323.2 K and 333.2 K and reference temperature ref$\_$t has been set accordingly. Pressure coupling is done using Berendsen barostat to pressure 1 bar. Isothermal compressibility of 4.6 x $10^{-5} \mbox{bar}^{-1}$ is used for box rescaling. Velocity is generated in agreement with thermal energy. During equilibration run all the bonds are converted into constraints using constraint algorithm LINCS.\\

		\noindent  After the equilibration, the system is ready for a production run. It is the step in which the diffusion coefficient is calculated. The simulation is done in NVT ensemble and there is no pressure coupling. All the bonds are held fixed using LINCS algorithm.

		\section*{Results and Discussion}
		\noindent In this section, we discuss the structural and dynamical properties of the constituents of the systems. 
		\section*{Structure of the System}
		\noindent Structural analysis of a system has been carried out by the help of radial distribution function (RDF). It gives the relative preference of particle's position in reference to certain atom as a function of radial distance. 
		\subsection*{Radial Distribution Function of Solvent }
		
		\noindent RDF of solvent reveals the equilibrium structure of the water molecules in the simulation box. As we have used SPC/E model, hydrogen in water does not take part explicitly in LJ interactions with any other atoms and its effects are included in united atom model. We have used RDF g$_{OW-OW}$ to study the structure of water molecules. 
		
		\begin{figure}[H]
			\includegraphics[scale=0.268]{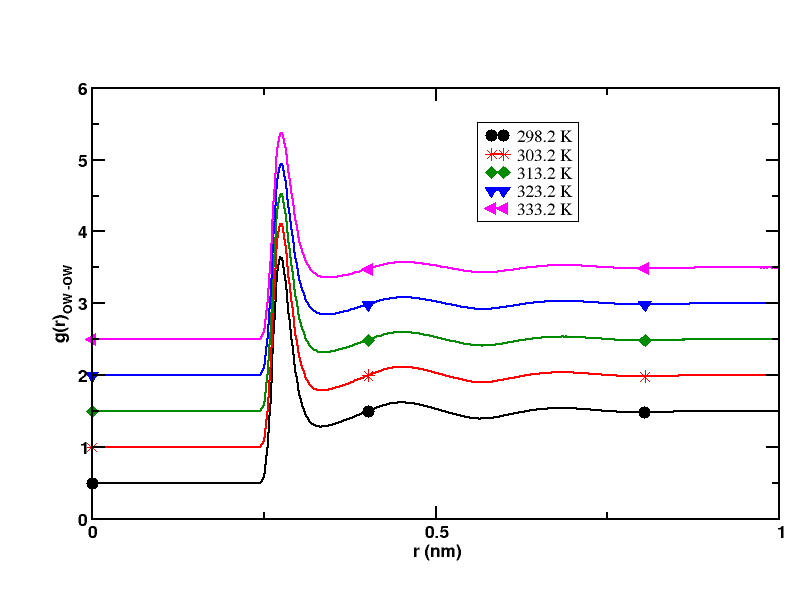}
			\caption{The RDF g$_{OW-OW}$ at different temperatures.}\label{rdf:OWOW}
		\end{figure}
		
		\noindent FIG. \ref{rdf:OWOW} represents the RDF g$_{OW-OW}$ at different temperatures. FIG. shows three distinct peaks of height 3.141, 1.126, and 1.046 at distance 0.274 nm, 0.450 nm, and 0.670 nm respectively at temperature 298.2 K. This explains that the radius of first coordination shell of oxygen of water around reference oxygen of water is 0.274 nm. Beyond the third peak, the plot is a straight line and has a mean value of one. With the increase in temperature, the peak height is decreasing while the width is increasing. As the temperature increases, the thermal agitation increases in temperature and this fact accounts for this effect in the graph. TABLE \ref{table:OWOW} shows the main result of RDF of a solvent. \\
		
		\begin{table}[H]
			\caption{Detail of RDF g$_{OW-OW}$ at different temperatures.}
			\label{table:OWOW}
			\resizebox{0.50\textwidth}{!}{\begin{tabular}{|c|c|c|c|c|c|c|c|}
					\hline
					Temperature (K) & ER(nm)  & FPP(nm) & FPV & SPP(nm) & SPV & THP (nm) & TPV\\
					\hline
					298.2     & 0.240 & 0.274 & 3.141 & 0.450 & 1.126 & 0.684 & 1.046\\
					\hline
					303.2      & 0.240 & 0.276 & 3.104 & 0.450 & 1.120 & 0.690 & 1.040\\
					\hline
					313.2  & 0.240 & 0.276 & 3.018 & 0.450 & 1.103 & 0.688 & 1.037\\
					\hline
					323.2     & 0.240 & 0.276 & 2.936 & 0.450 & 1.087 & 0.694 & 1.036\\
					\hline
					333.2     & 0.240 & 0.276 & 2.874 & 0.450 & 1.075 & 0.686 & 1.034\\
					\hline
					
				\end{tabular}}
			\end{table}

			\noindent The value of $\sigma$ for OW-OW is 0.3165 nm. The corresponding van der waals radius is 0.3553 nm. TABLE \ref{table:OWOW} shows that the value of excluded region is less than that of the van der Waals radius. This agrees with the theory that there is a zero probability of finding the particle in excluded region.
			
			\begin{figure}[H]
				\includegraphics[scale=0.3]{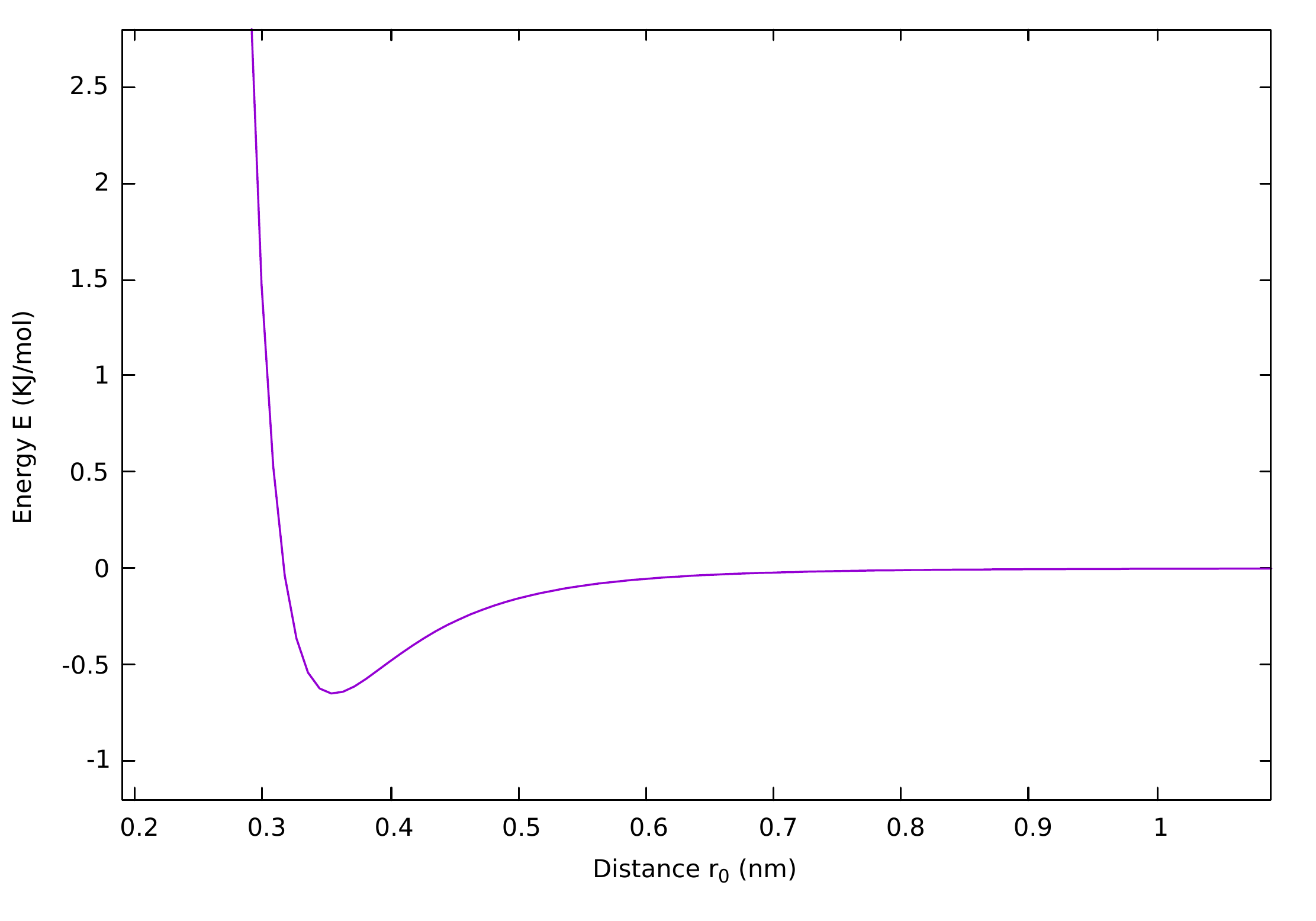}
				\caption{Lennard-Jones as a function of distance for two water molecules.}\label{lpowow}
			\end{figure}
			The FPP in FIG. \ref{table:OWOW} is nearly equal to the minimum of Lennard- Jones in FIG. \ref{lpowow}. The slightly less value of FPP in comparison to that of the minimum of Lennard -Jones potential indicates the many-body effects in the system~\cite{ipna}.

			\subsection*{Radial Distribution Function of OW-N1 }
\begin{figure}[H]
				\includegraphics[scale=0.293]{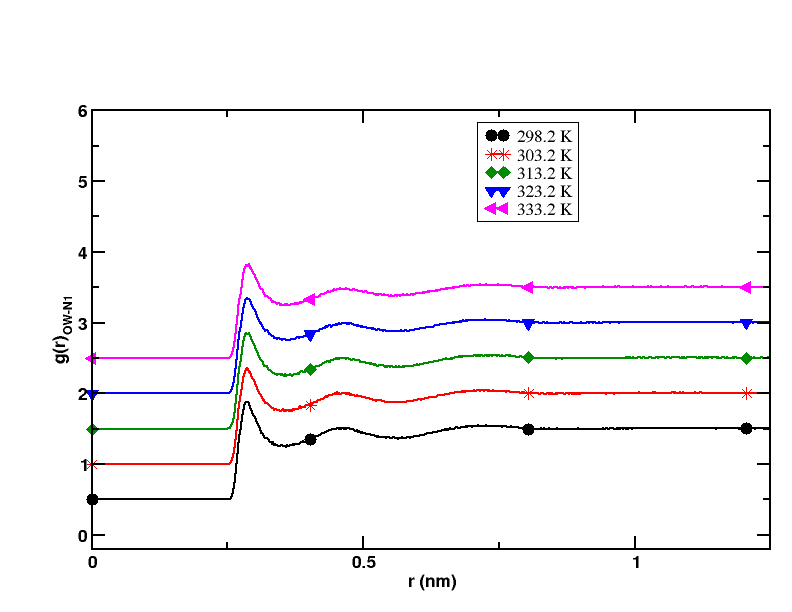}
				\caption{The RDF g$_{OW-N1}$ at different temperatures.}\label{rdf:OWN1}
			\end{figure}

			\noindent GABA contains one $NH_2$ group as discussed in chapter one. FIG.\ref{rdf:OWN1} depicts RDF g$_{OW-N1}$ at different temperatures. It gives probability of finding $N$ of GABA around the reference oxygen atom of water, relative to that for an ideal gas.

			\noindent FIG. \ref{rdf:OWN1} represents the RDF g$_{OW-N1}$ at different temperatures. \noindent At temperature 298.2 K, there are three peaks centered at 0.284 nm, 0.466 nm and 0.726 nm with peak height 1.379, 1.016,  and 1.047 respectively. At temperature 298.2 K, the most preferred position of oxygen in the water around N1 in GABA is at 0.284 nm. The RDF at 303.2 K, 313.2 K, 323.2 K and 333.2 K with peak position and peak height as presented in TABLE \ref{table:N1OW}. TABLE \ref{table:N1OW} shows the main result of FIG.\ref{rdf:OWN1} RDF of OW-N1 at different temperatures.\\
			
			\begin{table}[H]
				\caption{Detail of RDF g$_{OW-N1}$ at different temperatures.}
				\label{table:N1OW}
				
				\resizebox{0.50\textwidth}{!}{\begin{tabular}{|c|c|c|c|c|c|c|c|}
						\hline
						Temperature (K) & ER(nm) & FPP(nm) & FPV & SPP(nm) & SPV & THP (nm) & TPV\\
						\hline
						298.2     & 0.248 & 0.284 & 1.379 & 0.466 & 1.016 & 0.726 & 1.047\\
						\hline
						303.2 & 0.248 & 0.286 & 1.352 & 0.452 & 1.017& 0.712 & 1.047\\
						\hline
						313.2  & 0.246 & 0.284 & 1.359 & 0.470 & 1.004 & 0.730 & 1.048\\
						\hline
						323.2 & 0.248 & 0.286 & 1.344 & 0.468 & 0.997 & 0.724 & 1.045\\
						\hline
						333.2 & 0.248 & 0.290 & 1.321 & 0.458 & 0.985 & 0.726 & 1.040\\
						\hline
						
					\end{tabular}}
				\end{table}

				\noindent The value of $\sigma$ for OW-N1 is 0.3233 nm. The corresponding van der Waals radius is 0.3629 nm. TABLE \ref{table:N1OW} shows that the value of excluded region is less than that of the van der Waals radius. This agrees with the theory that there is a zero probability of finding the particle in the excluded region.

				\subsection*{Radial Distribution Function of OW-H7 }
\begin{figure}[H]
					\includegraphics[scale=0.293]{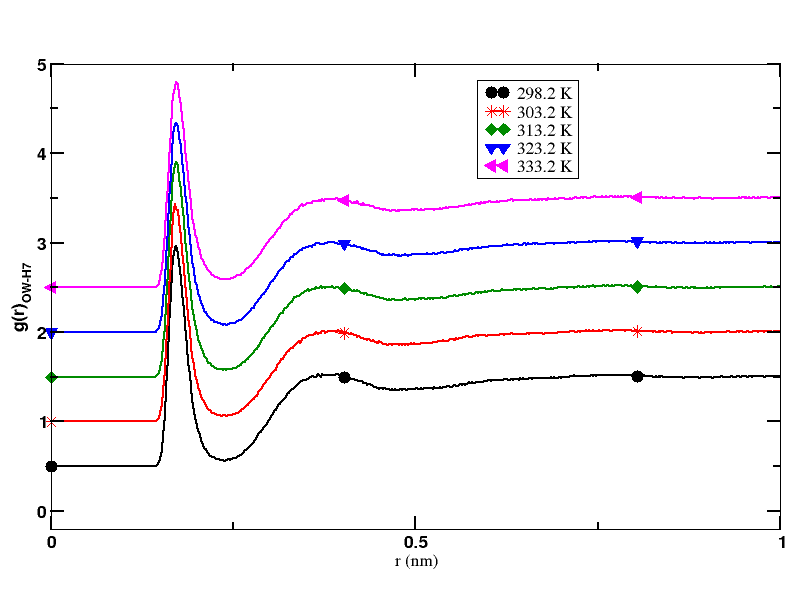}
					\caption{The RDF g$_{OW-H7}$ at different temperatures.}\label{rdf:OWH7}
				\end{figure}

				\noindent GABA contains one $NH_2$ group. FIG. \ref{rdf:OWH7} depicts RDF g$_{OW-H7}$ at different temperatures. H7 is the hydrogen of the  $NH_2$ group of GABA.
				TABLE \ref{table:OWH7} shows the main result of \ref{rdf:OWH7} RDF of OW -H7.

				\noindent FIG. \ref{rdf:OWH7} represents the RDF g$_{OW-H7}$ at different temperatures. 
				\begin{table}[H]
					\caption{Detail of RDF g$_{OW-H7}$ at different temperatures.}
					\label{table:OWH7}
					\resizebox{0.50\textwidth}{!}{
						\begin{tabular}{|c|c|c|c|c|c|c|c|}
							\hline
							Temperature (K) & ER(nm) & FPP(nm) & FPV & SPP(nm) & SPV & THP (nm) & TPV\\
							\hline
							298.2 & 0.138 & 0.172 & 2.467 & 0.372 & 1.030 & 0.780 & 1.024\\
							\hline
							303.2 & 0.142 & 0.170 & 2.431 & 0.394 & 1.024& 0.772 & 1.026\\
							\hline
							313.2  & 0.140 & 0.172 & 2.400 & 0.386 & 1.013 & 0.774 & 1.030\\
							\hline
							323.2 & 0.140 & 0.172 & 2.292 & 0.392 & 0.995 & 0.762 & 1.022\\
							\hline
							333.2 & 0.140 & 0.290 & 1.321 & 0.458 & 0.985 & 0.726 & 1.040\\
							\hline
							
						\end{tabular}}
					\end{table}  
					\noindent At temperature 298.2 K, there are three peaks centered at 0.172 nm, 0.372 nm and 0.780 nm with peak height 1.379, 1.016,  and 1.047 respectively. At temperature 298.2 K, the most preferred position of oxygen in the water around H7 in GABA is at 0.172 nm. The RDF at 303.2 K, 313.2 K, 323.2 K and 333.2 K with peak position and peak height as listed in TABLE \ref{table:N1OW}.\\
					
					\noindent The value of $\sigma$ for OW-H7 is 0.1583 nm. The corresponding van der Waals radius is 0.1776 nm. TABLE \ref{table:N1OW} shows that the value of excluded region is less than that of the van der Waals radius. This agrees with the theory that there is a zero probability of finding the particle in the excluded region.

					\subsection*{Radial Distribution Function of OW-O2 }
\begin{figure}[H]
						\includegraphics[scale=0.28]{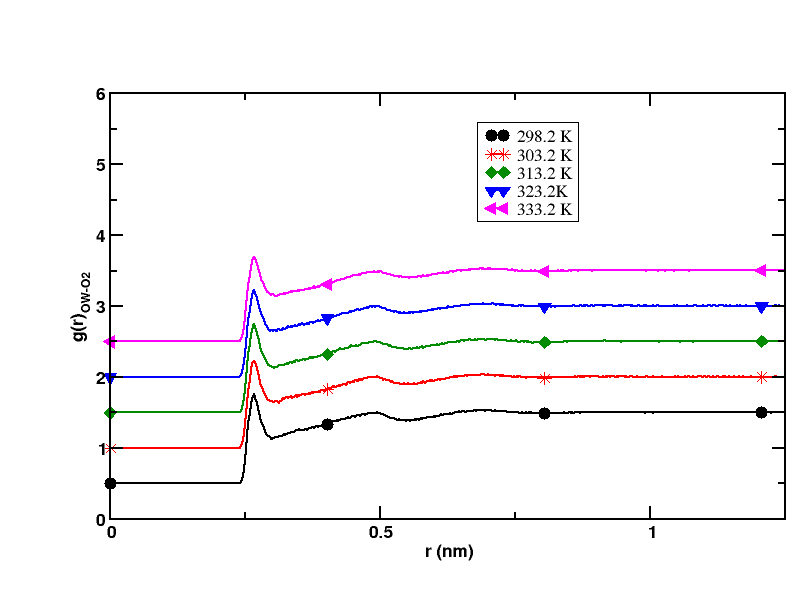}
						\caption{The RDF g$_{OW-O2}$ at different temperatures.}\label{rdf:OWO2}
					\end{figure}

					\noindent GABA contains one carboxylic group. FIG. \ref{rdf:OWO2} depicts RDF g$_{OW-O2}$ at different temperatures. O2 is the oxygen of the carboxylic group of GABA. 
					TABLE \ref{table:OWO2} shows the main result of FIG.\ref{rdf:OWO2} RDF of OW -H7.

					\noindent FIG. \ref{rdf:OWO2} represents the RDF g$_{OW-O2}$ at different temperatures. 
					
					\begin{table}[H]
						\caption{Detail of RDF g$_{OW-O2}$ at different temperatures.}
						\label{table:OWO2}
						
						\resizebox{0.50\textwidth}{!}{\begin{tabular}{|c|c|c|c|c|c|c|c|}
								\hline
								Temperature (K) & ER(nm)  & FPP(nm) & FPV & SPP(nm) & SPV & THP (nm) & TPV\\
								\hline
								298.2     & 0.234 & 0.266 & 1.258 & 0.494 & 1.003 & 0.700 & 1.039\\
								\hline
								303.2  & 0.236 & 0.266 & 1.231 & 0.494 & 1.008& 0.700 & 1.039\\
								\hline
								313.2 & 0.234 & 0.266 & 1.242 & 0.490 & 1.006& 0.684 & 1.034\\
								\hline
								323.2 & 0.234 & 0.266 & 1.228 & 0.494 & 0.999& 0.690 & 1.037\\
								\hline
								333.2 & 0.234 & 0.266 & 1.188 & 0.488 & 0.991& 0.684 & 1.031\\
								\hline
								
							\end{tabular}}
						\end{table}

						\noindent At temperature 298.2 K, there are three peaks centered at 0.266 nm, 0.494 nm and 0.704 nm with peak height 1.258, 1.003,  and 1.039 respectively. At temperature 298.2 K, the most preferred position of oxygen in the water around O2 in GABA is at 0.266 nm. The RDF at 303.2 K, 313.2 K, 323.2 K and 333.2 K with peak position and peak height as listed in TABLE \ref{table:OWO2}.\\
						
						\noindent The value of $\sigma$ for OW-O2 is 0.3083 nm. The corresponding van der Waals radius is 0.3461 nm. TABLE \ref{table:OWO2} shows that the value of excluded region is less than that of the van der Waals radius. This agrees with the theory that there is a zero probability of finding the particle in the excluded region.
						
						\section*{Diffusion Coefficients}
						
						\subsection*{Self Diffusion Coefficient of GABA}
						
						\noindent We have estimated self diffusion coefficient $D_G^S$ of GABA at different temperatures. The self diffusion coefficient of GABA was found to be increased with increase in temperature. 
						
						\begin{figure}[H]
							\centering
							\includegraphics[scale=0.293]{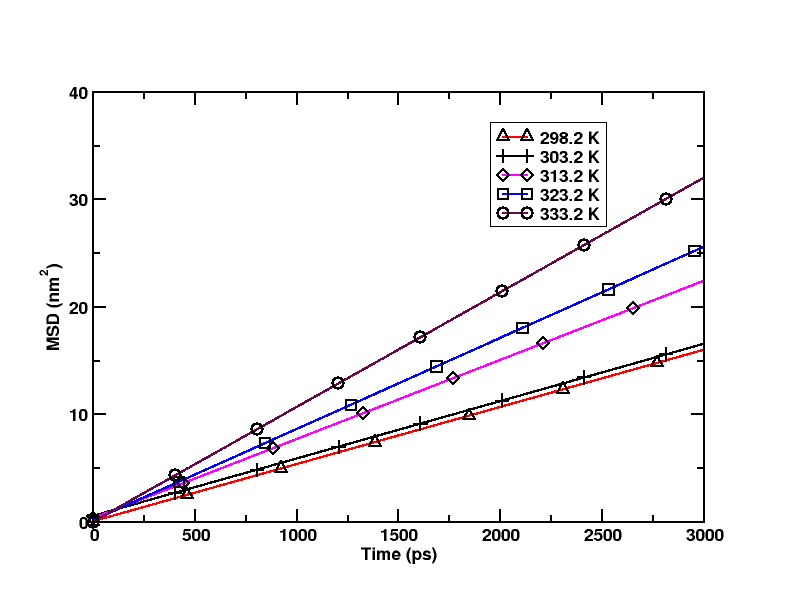}
							\caption{MSD plots for GABA obtained from simulation and their linear fit at different temperatures.}
							\label{msdgabaall}
						\end{figure}
						\noindent FIG. \ref{msdgabaall} shows the MSD plots of GABA obtained from the simulation for 3 ns and their respective linear fit at different temperatures. As seen in FIG.\ref{msdgabaall}, the slope of MSD plot increases with the increase in temperature. Self - diffusion coefficient of GABA also increases with the increase in temperature.
						
						\subsection*{Self Diffusion coefficient of Water}
						
						We have estimated self diffusion coefficient $D_W^S$ of water at different temperatures. The self diffusion coefficient of water was found to be increased with increase in temperature.

						\begin{figure}[H]
							\centering
							\includegraphics[scale=0.283]{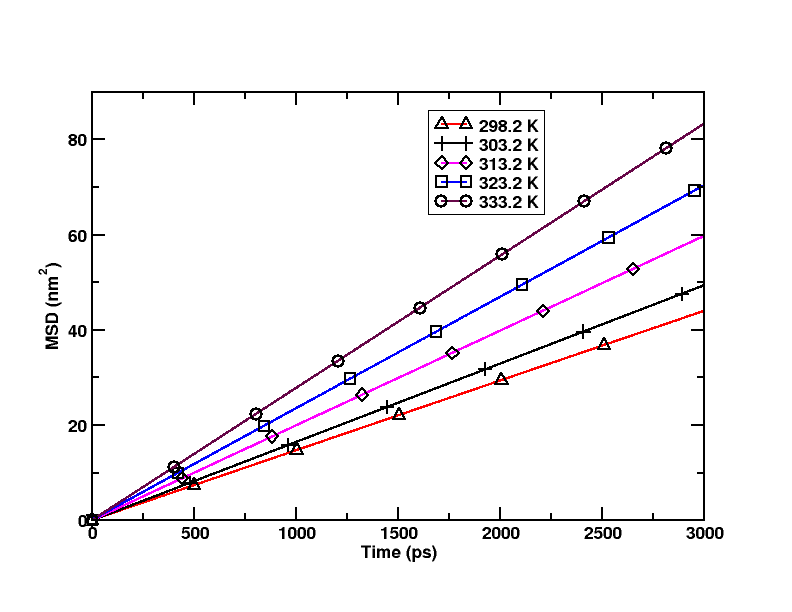}
							\caption{MSD plot of water at different temperature.}
							\label{msdwaterall}
						\end{figure}
						\noindent FIG. \ref{msdwaterall} depicts the MSD plots of water obtained from the simulation for 3 ns and their corresponding linear fit at different temperatures. From FIG.\ref{msdwaterall} it is clear that slope of the MSD plot increases with the increase in temperature. As the self - diffusion coefficient is directly related with a slope of MSD plot, we can conclude that the self - diffusion coefficient of water increases with the increase in temperature.

						\subsection*{Binary Diffusion Coefficient of GABA-water System}
						\noindent Our system has 3 GABA molecules and 1035 water molecules for 298.2 K, 323.2 K and 333.2 K. Similarly, for temperatures 303.2K and 313.2 K, our system has 3 GABA molecules and 1050 water molecules. The mole fraction of GABA is 0.0028 and  that of water is 0.997. The mole fraction of GABA is much smaller in comparison to that of water, thus, the term containing $N_W$ dominates the term containing $N_G$ in Darken's relation. 
						
						The binary-diffusion coefficient of GABA obtained from simulations are presented and compared in the TABLE \ref{table:binaryDiff}:\\
						
						\begin{table}[H]
							\caption{Binary - diffusion coefficient of GABA in water at different temperatures}\label{table:binaryDiff}
							\resizebox{0.50\textwidth}{!}{
								\begin{tabular}{|c|c|c|c|c|}
									\hline
									SN & Temperature & $D_{GW}^S$ ( $10^{-10} m^2/s$) & $D_{GW}^E$ ( $10^{-10} m^2/s$) \cite{kyui} & Error (\%) \\
									
									\hline
									1. & 298.2 & 8.81  & 8.38 & 5.13\\
									\hline
									2. & 303.2  &   9.87&9.54  & 3.45\\
									\hline
									3. & 313.2  &  12.81 & 11.99 & 6.83\\
									\hline
									4. & 323.2  &  14.22 & 14.89  & 4.49\\
									\hline 
									5. & 333.2 & 17.73& 17.89 & 0.89\\
									\hline
									
								\end{tabular}
							}
						\end{table}
						
						\noindent TABLE \ref{table:binaryDiff} presents the binary - diffusion coefficient of GABA - water mixture obtained from the simulation at different temperatures and comparison of simulated results with available experimental results. The binary - diffusion coefficients of our system at different temperatures obtained from MD simulations are in good agreement with experimental results within the error of $7\%$.

						\begin{figure}[H]
							\includegraphics[scale=0.31]{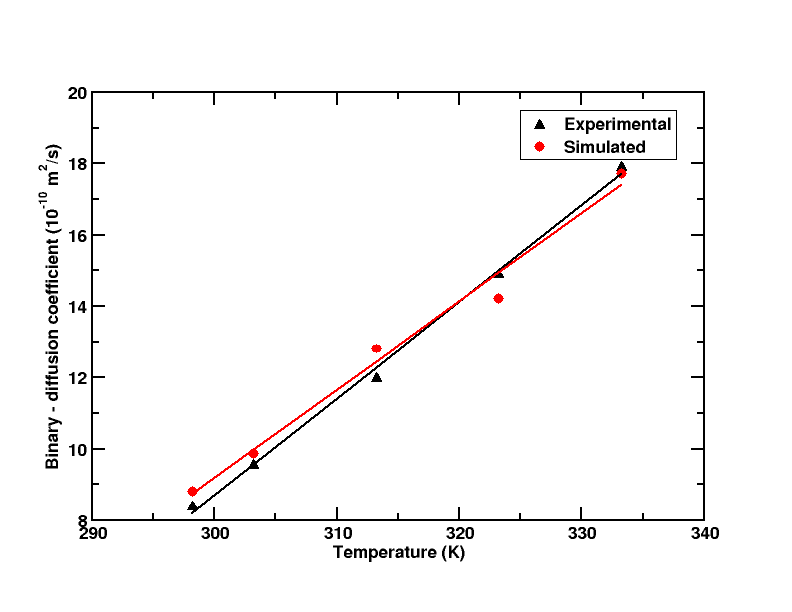}
							\caption{Binary-diffusion coefficient of GABA in water at different temperature obtained from simulation and from literature \cite{kyui}}
							\label{fig:binaryDiff}
						\end{figure}
						
						\noindent FIG. \ref{fig:binaryDiff} is the graphical representation of simulated and experimental result for the binary-diffusion coefficient of GABA in water at a different temperatures. The binary diffusion coefficient of GABA-water mixture at different temperatures is found to be in good agreement with the experimental data reported in \cite{kyui}
						
						\section*{Temperature Dependency of Diffusion}
						Temperature is one of the important factors for diffusion. The diffusion is high at higher temperature whereas low at a lower temperature. The relation between the diffusion and temperature is given by the Arrhenius formula~\cite{crank}.
						
						\begin{equation}
						\label{arrhenius}
						\ln D = \ln D_o - \frac{E_a}{N_A k_B T}
						\end{equation}
						\noindent Here. D$_o$ is called pre-exponent factor,
						E$_a$ is the activation energy for diffusion,
						T is the absoulute temperature, 
						N$_A$ is Avogadro number, 
						k$_B$ is the Boltzmann constant
						
						The Arrhenius plot is plot between ln(\textit{D}) and reciprocal of absolute temperature (T). The slope of the Arrhenius plot gives the activation energy of the diffusion process \cite{crank}.\\
						The activation energy is given by,
						\begin{equation}
						\label{activationenergy}
						E_a = - N_Ak_B \frac{\partial\ln (D)}{\partial (1/T)}
						\end{equation}
						\noindent The intercept, extrapolated to the 1/T $\rightarrow 0$, gives the pre-exponential factor.
						
						\begin{figure}[H]
							\centering
							\includegraphics[scale=0.293]{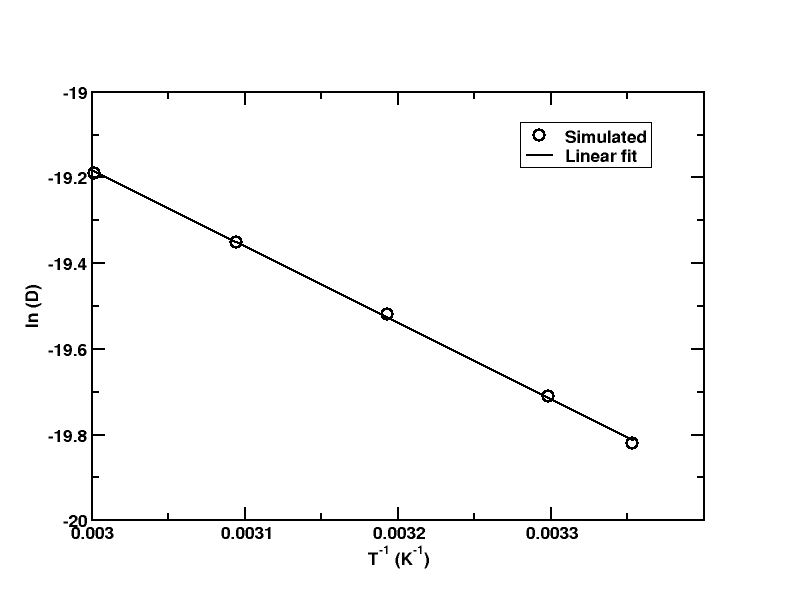}
							\caption{Arrhenius diagram of the self-diffusion coefficient of water.}
							\label{arhw}
						\end{figure}
						\noindent FIG.\ref{arhw} shows the Arrhenius diagram for self - diffusion coefficient of water. The activation energy for self - diffusion coefficient of water estimated from the slope is equal to 14.77 kJ mol$^{-1}$.\\
						\begin{figure}[H]
							\centering
							\includegraphics[scale=0.293]{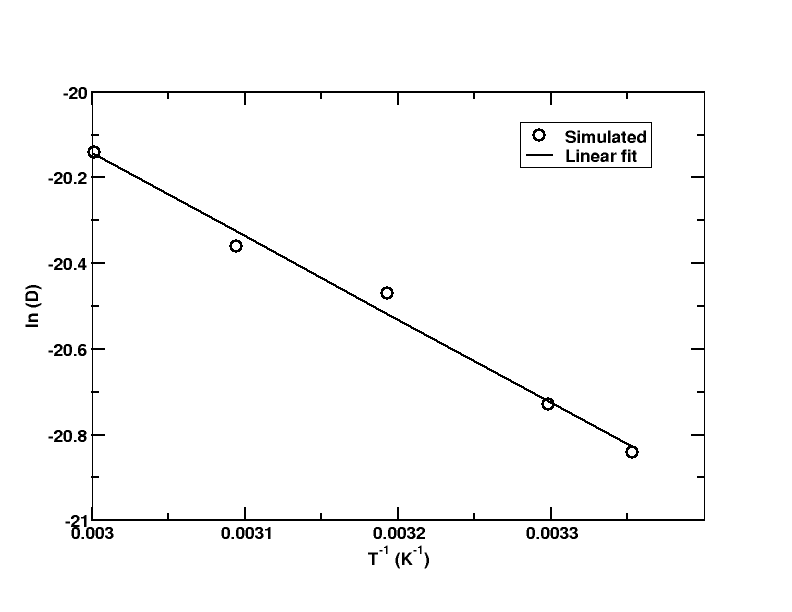}
							\caption{Arrhenius diagram of the self-diffusion coefficient of GABA.}
							\label{arhg}
						\end{figure}
						\noindent FIG.\ref{arhg} shows the Arrhenius diagram for self - diffusion coefficient of GABA. The activation energy for self - diffusion coefficient of GABA estimated from the slope is equal to  16.13 kJ mol$^{-1}$.\\
						
						\begin{figure}[H]
							\centering
							\includegraphics[scale=0.293]{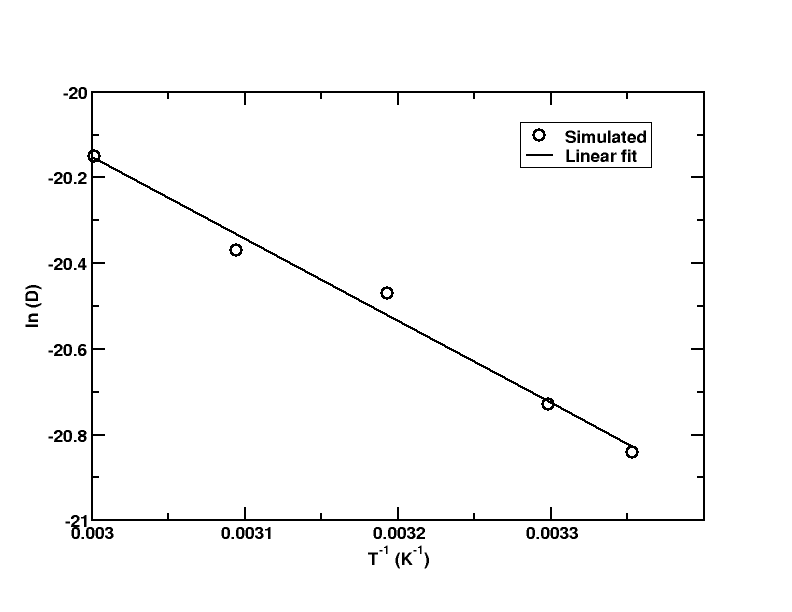}
							\caption{Arrhenius diagram of the binary - diffusion coefficient of GABA.}
							\label{arhb}
						\end{figure}
						\noindent FIG.\ref{arhb} shows the Arrhenius diagram for binary - diffusion coefficient of water - GABA system. The activation energy for binary - diffusion coefficient of water - GABA estimated from the slope is equal to 15.85 J mol$^{-1}$.\\
						
						\noindent With the help of Arrhenius diagrams, we can estimate the activation energy of our system. The activation energy as estimated from simulation  are presented in TABLE \ref{table:activationenergy}. The activation energy of water was compared with the experimental results ~\cite{easteal}. The estimated value of activation energy of water agreed well with the experimental value within an error of around 17 \%. This error can be accounted to the fact that our system contains both the GABA molecules and water. But the activation energy reported in the experiment~\cite{easteal} is of pure water only.

						\begin{table}[H]
							\centering
							\caption{Activation Energies for Diffusion.}
							\label{table:activationenergy}
							\begin{tabular}{|c|c|c|c|}
								\hline
								System  & \multicolumn{2}{|c|}  {Activation energy (E$_a$) in kJ mol$^{-1}$} & Error Percentage\\
								\hline
								& Simulated & Experimental & \\
								\hline    
								Water  & 14.77 &  17.79 ~\cite{easteal} & 16.95 \%\\
								\hline
								GABA & 16.13  & &  \\
								\hline
								Binary mixture  &15.85 &  & \\
								\hline
							\end{tabular}
						\end{table}
						
						\section*{Conclusions and Concluding Remarks}
						\noindent A molecular dynamics study of a system containing water molecules and GABA molecules were carried out at temperatures 298.2 K, 303.2 K, 313.2 K, 323.2 K and 333.2 K. The simulation was done using software package GROMACS. The solute and solvent are modeled using OPLS-AA force field. The equilibration was done for 200 ns and the production run was done for 100 ns. PME was used for coulomb interaction, as a result of which there was no restriction in Coulomb interaction.\\
						
						\noindent The energy profile at different temperature are studied to check the equilibration of our system. At 298.2 K, the average total energy  is -40612.70 $\pm$ 0.38 kJ mol$^{-1}$. Since the total energy is negative, the system at 298.2 K is also bound and stable. Similar results were obtained for other temperatures too. The transfer of mass, diffusion of the system is studied. Self diffusion coefficients of GABA and water are studied using Einstein's method. The self diffusion coefficients are calculated using MSD plots. The binary diffusion coefficient was calculated using the Darken's relation. The binary diffusion coefficient of GABA-water mixture at different temperatures is found to be in good agreement with the experimental data reported in \cite{kyui} within error $7\%$. RDF of different pairs of atoms OW-OW, OW-H7, OW-O2 and OW-N1 are used to study the equilibrium structure of the system. The diffusion coefficients were found to be dependent on the temperature as they showed Arrhenius behavior. We can conclude that molecular dynamics is an effective tool for the study of transport properties of biomolecules. The calculated values of diffusion coefficient at different temperatures can be taken as reference for future studies. \\

						\noindent This study can be extended in future to the study of diffusion of GABA in another solvents like methanol, acetonitrile, etc. Moreover, other transport properties such as thermal conductivity and viscosity can also be studied. We can also carry out similar work for other amino acids.
						
						\section*{Acknowledgements}
						\noindent EM acknowledges the financial support from University Grants Commission (UGC), Nepal.

					\end{document}